\documentclass[conference]{IEEEtran}

\usepackage[pdftex]{graphicx}
\usepackage{url}
\usepackage{caption}

\begin{document}
	
	\renewcommand{\thetable}{\arabic{table}}
	\renewcommand{\figurename}{Fig.}

\title{Cyber-Deception and Attribution\\ in Capture-the-Flag Exercises}

\author{\IEEEauthorblockN{Eric Nunes, Nimish Kulkarni, Paulo Shakarian}
	\IEEEauthorblockA{School of Computing, Informatics and\\Decision Systems Engineering\\
		Arizona State University\\
		Tempe, AZ 85281, USA\\
		Email: \{enunes1, nimish.kulkarni, shak\} @asu.edu}
	\and
	\IEEEauthorblockN{Andrew Ruef, Jay Little }
	\IEEEauthorblockA{Trail of Bits, Inc.\\
		New York, NY 10003, USA\\
		Email: \{andrew, jay\} @trailofbits.com}
	}

\maketitle

\begin{abstract}
Attributing the culprit of a cyber-attack is widely considered one of the major technical and policy challenges of cyber-security. The lack of ground truth for an individual responsible for a given attack has limited previous studies. Here, we overcome this limitation by leveraging DEFCON capture-the-flag (CTF) exercise data where the actual ground-truth is known. In this work, we use various classification techniques to identify the culprit in a cyberattack and find that deceptive activities account for the majority of misclassified samples. We also explore several heuristics to alleviate some of the misclassification caused by deception.
\end{abstract}

\IEEEpeerreviewmaketitle

\section{Introduction}
Attributing the culprit of a cyber-attack is widely considered one of the major technical and policy challenges of cyber-security. The lack of ground truth for an individual responsible for a given attack has limited previous studies. In this study, we take an important first step toward developing computational techniques toward attributing the actual culprit (here hacking group) responsible for a given cyber-attack. We leverage DEFCON capture-the-flag (CTF) exercise data which
we have processed to be amenable to various machine learning approaches. Here, we use various classification techniques to identify the culprit in a cyber-attack and find that deceptive
activities account for the majority of misclassified samples. We also explore several heuristics to alleviate some of the misclassification caused by deception. Our specific contributions are as follows:

\begin{itemize}
	\item We assemble a dataset of cyber-attacks with ground truth derived from the traffic of the CTF held at DEFCON 21 in 2013.
	\item We analyze this dataset to identify cyber-attacks where deception occurred.
	\item We frame cyber-attribution as a multi-label classification problem and leverage several machine learning approaches. We find that deceptive incidents account for the vast majority of misclassified samples.
	\item We introduce several pruning techniques and show that they can reduce the effect of deception as well as provide insight into the conditions in which deception was employed by the participants of the CTF.
\end{itemize}

In our text on cyber-warfare~\cite{warfare}, we discuss the difficulties of cyber-attribution and how an intelligence analyst must also explore the deception hypothesis in a cyber-warfare scenario.  When compared to other domains of warfare, there is a much greater potential for evidence found in the aftermath of cyber-attack to be planted by the adversary for purposes of deception.  The policy implications of cyber-attribution have also been discussed in~\cite{policy} where the authors point out that anonymity, ability to launch multi-stage attacks, and attack speed pose significant challenges to cyber attribution.

In an early survey on cyber-attribution~\cite{survey}, the authors point out that technical attribution will generally identify machines, as opposed to a given hacker and his/her affiliations.  While we will use technical information in our approach, we have ground truth data on the group involved by the nature of the capture-the-flag data.  This will allow our approach to profile the tactics, techniques, and procedures of a given group as we have ground-truth information on a hacking group as opposed to machines.  An example of such an approach is the WOMBAT attribution method~\cite{wombat} which attributes behavior to IP sources that are potentially linked to some root cause determined through a clustering technique.  Similarly, other work~\cite{Thonnard} combines cluster analysis with a component for multi-criteria decision analysis and studied an implementation of this approach using honeypot data -- again, this approach lacks any ground truth of the actual hacker or hacking group.

Concurrently,  we have devised a formal logical framework for reasoning about cyber-attribution~\cite{shakarian,shakarian14}.  However, we have not studied how this framework can be instantiated on a real world dataset and, to date, we have not reported on an implementation or experiments in the literature. We note that none of the previous work on cyber-attribution leverages a data set with ground truth information of actual hacker groups -- which is the main novelty of this paper.

\section{Dataset}
\label{data}
Our dataset consists of events recorded from a Capture-the-flag (CTF) tournament held at DEFCON 21 in 2013. Briefly, CTF competitions act as educational exercise that exposes real world attack scenarios to participants. Network sniffing, analysis of protocols, programming and system level knowledge, cryptanalysis are some of the instrumental skills acquired by contestants. 

 Our data represents attack/defense style, where each team owns a small network of machines to defend. Teams are judged based on scores given to attack machines of other teams as well as defending their own network. Initially, all virtual machines are configured with specific set of services. These services are vulnerable to state-of-art hacking techniques. Files can be considered as form of flag to be captured from other teams or to be planted to other teams by exploiting those vulnerabilities.

DEFCON CTF organizers recorded network traffic that includes network packets generating to and from all participating teams and is available on Internet~\cite{pcap}. Recordings are stored as archive files of PCAP (packet capture) for each team (destination as that team) separately. PCAP file contains packet headers (TCP, SSL, UDP etc.) and respective data as source, destination, sequence numbers etc. with timestamp having millisecond precision. Using open source tool tcpflow\footnote{https://github.com/simsong/tcpflow}, we interpreted collection of PCAPs as cumulative data streams. Tcpflow reconstructs actual data streams from the packets that proved helpful in protocol analysis and debugging. 
This tool produces a file containing the contents of each stream, representing the data sent between two points in the CTF system. 

For each file, we computed an md5 checksum, a byte histogram, and an ARM instruction histogram. This data was recorded as a list of tuples (time-stamp, hash, byte-histogram, instruction-histogram) in a JSON document. These individual fields of the tuple are listed in Table~\ref{exAt}. 
\vspace{-1em}
\begin{table}[h!]
	\caption{\textmd{Fields in an instance of network attack}}
	\label{exAt}
	\tiny
	\centering
	\renewcommand{\arraystretch}{1.5}
	
	\begin{tabular}{|p{2.3cm}|p{4cm}|} 
		\hline
		{\bf Field} &  {\bf Intuition} \\ \hline 
		
		\textsf{byte\_hist} & histogram of byte sequences in the payload\\ \hline
		\textsf{inst\_hist} & histogram of instructions used in the payload\\ \hline
		\textsf{from\_team} & the team where the payload originates (attacking team)\\ \hline
		\textsf{to\_team} & the team being attacked by the payload\\ \hline
		\textsf{svc} & the service that the payload is running\\ \hline
		\textsf{payload\_hash} & indicates the payload used in the attack (md5)\\ \hline
		\textsf{time} & indicates the date and time of the attack\\
		\hline
	\end{tabular}
	\vspace{-1em}
	
\end{table}
\vspace{-0.5em}

From this pre-processing of the network data (packets) we have around 10 million network attacks. There are 20 teams in the CTF competition. In order to attribute an attack to a particular team, apart from analyzing the payloads used by the team, we also need to analyze the behavior of the attacking team towards his adversary. For this purpose we separate the network attacks according to the team being targeted. Thus we have 20 such subsets. We represent the 20 subsets (teams) as T-i, where i = 1, 2, 3...20. An example of an event in the dataset is shown in Table~\ref{event}.
\vspace{-1em}
\begin{table}[h!]
	\caption{\textmd{Example event from the dataset}}
	\label{event}
	\tiny
	\centering
	\renewcommand{\arraystretch}{1.5}
	
	\begin{tabular}{|p{2.3cm}|p{4cm}|} 
		\hline
		{\bf Field} &  {\bf Value} \\ \hline 
		
		\textsf{byte\_hist} & $0{\times}43$:245, $0{\times}69$:8, $0{\times}3a$:9, $0{\times}5d$:1, .....\\ \hline
		\textsf{inst\_hist} & cmp:12 , svcmi:2, subs:8, movtmi:60 ......\\ \hline
		\textsf{from\_team} &  men in black hats\\ \hline
		\textsf{to\_team} & Robot Mafia\\ \hline
		\textsf{svc} & 02345\\ \hline
		\textsf{payload\_hash} & 2cc03b4e0053cde24400bbd80890446c\\ \hline
		\textsf{time} & 2013-08-03T23:45:17\\
		\hline
	\end{tabular}
	\vspace{-1em}
	
\end{table}

\subsection{Dataset Analysis}
\begin{figure}[htp!]
	\centerline{\includegraphics[scale=0.24,keepaspectratio]{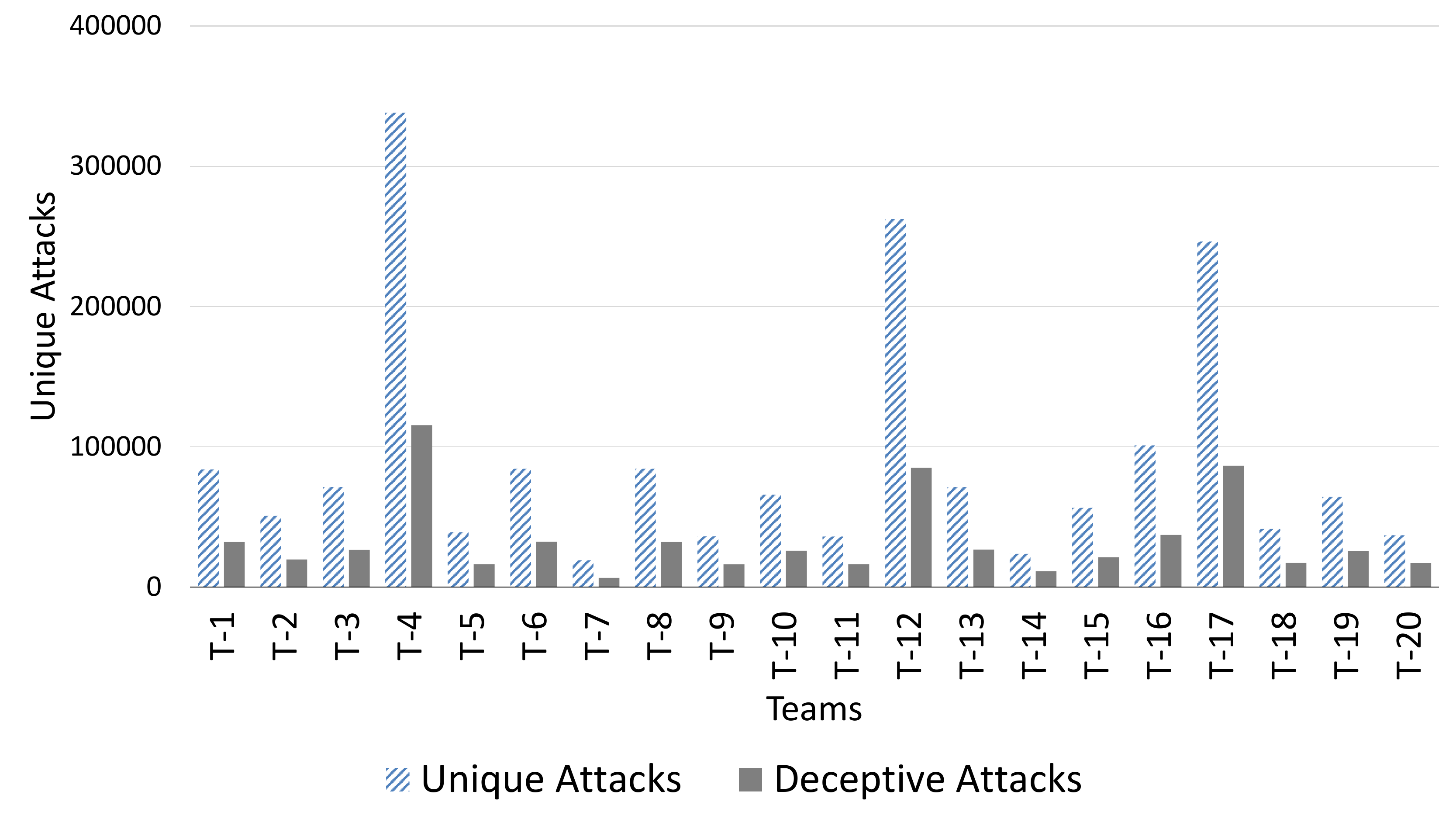}}
	\captionsetup{justification=centering}
	
	\caption{\textmd{Unique deceptive attacks directed towards each team.}}

	\label{fig:one}
\end{figure}

We now discuss two important observations from the dataset, that makes the task of attributing an observed network attack to a team difficult.\smallskip \\
\noindent\textit{Deception:}
In the context of this paper we define an attack to be deceptive when multiple adversaries get mapped to a single attack pattern. In the current setting we define deception as the scenario when the same payload is used by multiple teams to target the same team. \figurename~\ref{fig:one} shows the distribution of unique deception attacks with respect to the total unique attacks in the dataset based on the target team. These unique deceptive attacks amount to just under 35\% of the total unique attacks. 

\begin{figure}[htp!]
	\centerline{\includegraphics[scale=0.24,keepaspectratio]{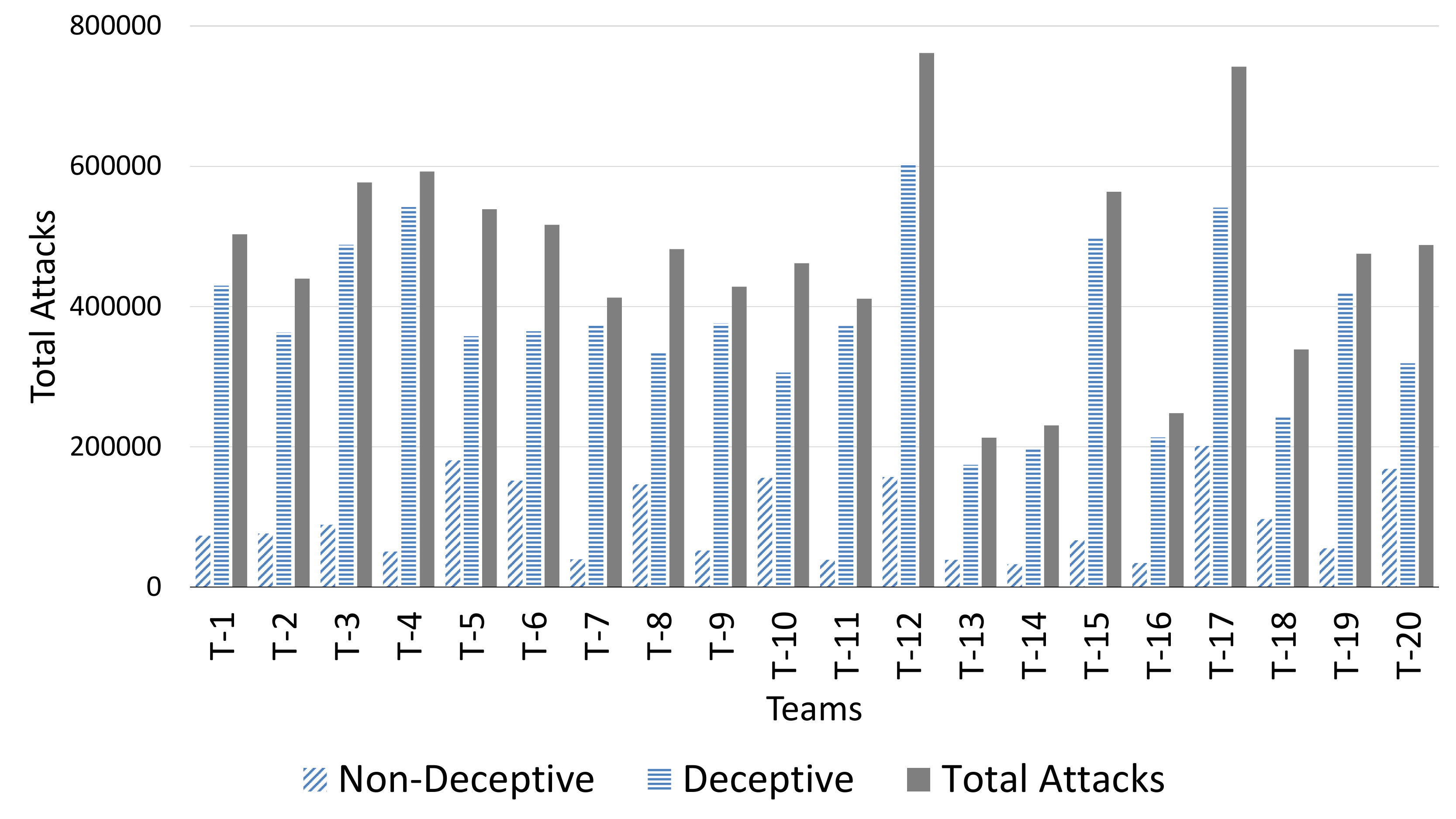}}
	\captionsetup{justification=centering}
	
	\caption{\textmd{Total attacks and duplicate attacks(Deceptive and Non-deceptive) directed towards each team}}
	
	\label{fig:two}
\end{figure}

\vspace{-1em}
\noindent\textit{Duplicate attacks:}
A duplicate attack occurs when the same team uses the same payload to attack a team at different time instances. Duplicate attacks can be attributed to two reasons. First when a team is trying to compromise other systems, it just does not launch a single attack but a wave of attacks with very little time difference between consecutive attacks. Second, once a successful payload is created which can penetrate the defense of other systems, it is used more by the original attacker as well as the deceptive one as compared to other payloads. We group duplicates as being non-deceptive and deceptive. Non-deceptive duplicate are the duplicates of the team that first initiated the use of a particular payload. On the other hand deceptive duplicates are all the attacks from the teams that are being deceptive. Deceptive duplicates form a large portion of the dataset as seen in \figurename~\ref{fig:two}.

\section{Baseline Approaches}
\label{baselineAppr}
From the dataset, we have the ground truth available for all the samples. Hence we use supervised machine learning approaches to predict the attacking team. The ground truth corresponds to a team competing in the competition.

\noindent\textbf{Decision Tree (DT).}  For baseline comparisons we first implemented a decision tree classifier. We built the decision tree by finding the attribute that maximizes the information gain at each split. In order to avoid over-fitting, the terminating criteria is set to less than 0.1\% of total samples.\smallskip \\
\noindent\textbf{Random Forest (RF).}  We use a random forest which combines bagging for each tree with random feature selection at each node to split the data thus generating multiple decision tree classifiers. \smallskip\\
\noindent\textbf{Support Vector Machine (SVM).}  Support vector machines is a popular supervised classification technique that works by finding a separating margin that maximizes the geometric distance between classes. We use the popular LibSVM implementation~\cite{Chang2011} which is publicly available. \smallskip \\
\noindent\textbf{Logistic Regression (LOG-REG).}  Logistic regression classifies samples by computing the odds ratio. The odds ratio gives the strength of association between the features and the class. We implement the multinomial logistic regression which handles multi-class classification.

\subsection{Experimental Results}
\label{results}
\begin{figure}[htp!]
	\centerline{\includegraphics[scale=0.24,keepaspectratio]{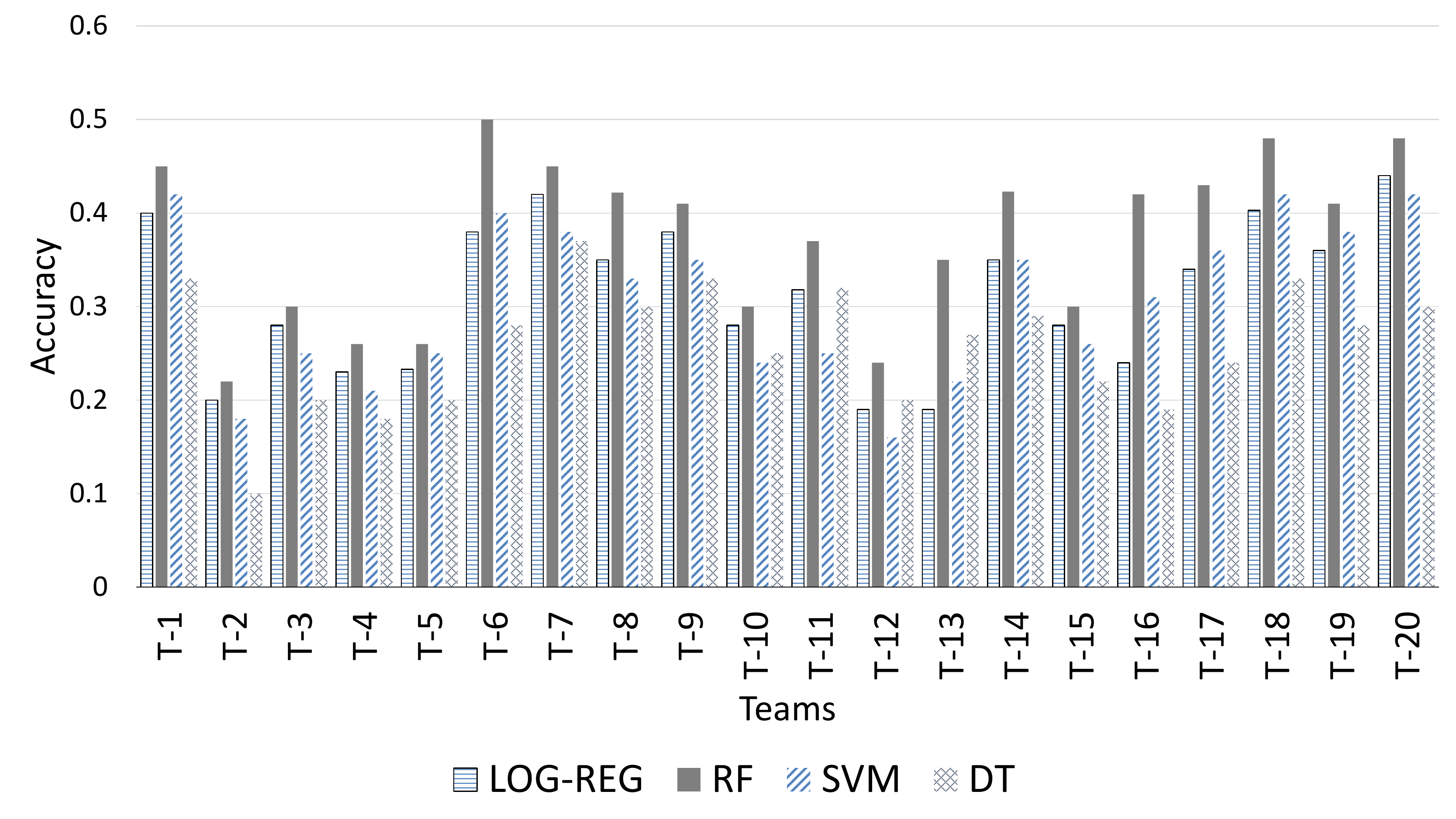}}
	\captionsetup{justification=centering}
	\caption{\textmd{Team prediction accuracy for LOG-REG, RF, SVM and DT.}}
\label{fig:three}
\end{figure}

For our baseline experiments, we separate the attacks based on the team being targeted. Thus we have 20 subsets. We then sort the attack according to time. We reserve the first 90\% of the attacks for training and the rest 10\% for testing. Attacker prediction accuracy is used as the performance measure for the experiment. Accuracy is defined as the fraction of correctly classified test samples. \figurename~\ref{fig:three} shows the accuracy for predicting the attacker for each target team. Machine learning techniques significantly outperform random guessing which would have an average accuracy of choosing 1 out of 19 teams attacking yielding an accuracy of 0.053. For this experiment random  forest classifier performs better than logistic regression, support vector machine and decision tree for all the target teams. Table~\ref{acc} below summarizes the average performance for each method.

\begin{table}[h!]
	\caption{\textmd{Summary of Prediction results averaged across all Teams}}
	\label{acc}
	\centering
	\tiny
	\renewcommand{\arraystretch}{1.5}
	
	\begin{tabular}{|p{4cm}|p{2cm}|} 
		\hline
		{\bf Method} &  {\bf Average Performance} \\ \hline \hline
		\textsf{Decision tree (DT)} & 0.26\\ \hline
		\textsf{Logistic regression (LOG-REG)} & 0.31\\ \hline
		\textsf{Support vector machine (SVM)} & 0.30\\ \hline
		\textsf{Random Forest (RF)} & \textbf{0.37}\\ \hline

	\end{tabular}
	\vspace{-1em}
	
\end{table}

\subsection{Misclassified Samples}	
Misclassification can be attributed to the following sources,
\begin{itemize}
	\item Non-deceptive duplicate attacks attributed to one of the deceptive teams. 
	\item Deceptive duplicates attributed to some other deceptive team.
	\item Payloads that were not encountered during the training phase.
\end{itemize}
The first two sources of error make up the majority of misclassifications, since a given attack can be attributed to any of the 19 teams.  
	
\begin{figure}[htp!]
	\centerline{\includegraphics[scale=0.24,keepaspectratio]{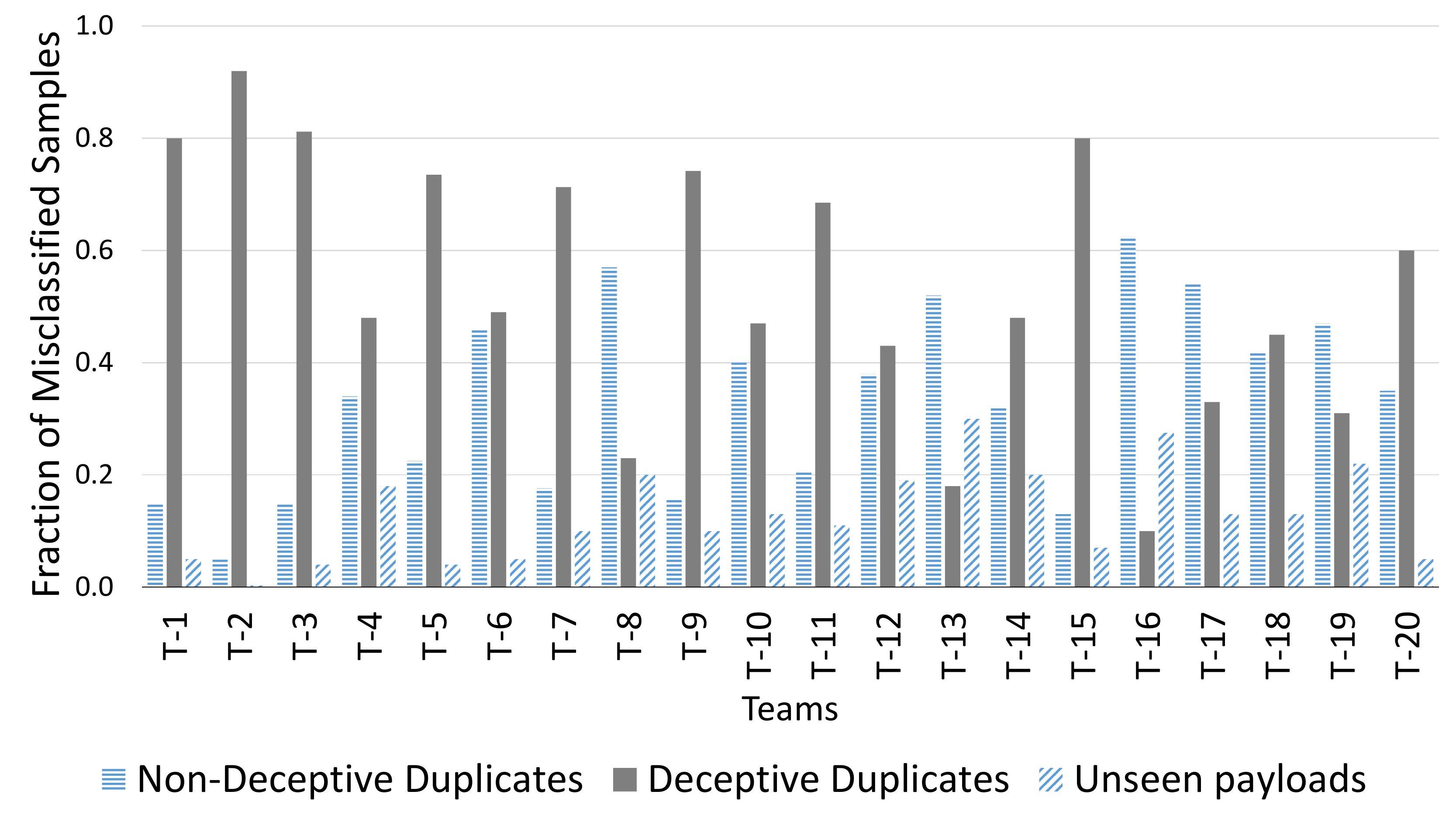}}
	\captionsetup{justification=centering}
	
	\caption{\textmd{Sources of error in the misclassified samples.}}
	
	\label{fig:four}
\end{figure}
\figurename~\ref{fig:four} shows the distribution of the above mentioned sources of misclassification for each team. Deceptive duplicates form the majority of misclassifications. This is not surprising given the fact that deceptive duplicates make up almost 90\% of the total attacks (see \figurename~\ref{fig:two}).

\section{Pruning}
\label{prun}
We explore different pruning techniques to address misclassification issues with respect to deceptive and non-deceptive duplicates. The pruning techniques are only applied to the training data, while the test data is maintained at 10\% as mentioned in Section~\ref{results}. We use the random forest classifier for all the pruning techniques.\\

\noindent\textbf{All-but-majority (P-1):}
In this pruning technique, for each payload, we only retain duplicates of the most frequent attacking team and prune the duplicates of all other teams. This pruned set is then used to train the random forest classifier. Table~\ref{teams} shows the classifier performance in comparison with the baseline method. All-but-majority pruning technique has better performance on the test set than the baseline approach for 11 out of 20 teams. Using this pruning technique does benefit majority of the teams as the prediction accuracy improves for them, but for some teams the performance drops. The reason for the drop in performance for some teams is due to the fact that training set gets dominated by a single team which does not have majority in testing set. Since the majority team gets represented in most of the leaves of the random forest classifier, it gets predicted more often leading to high misclassifications.\smallskip \\
\noindent\textbf{All-but-K-majority (P-2):}
In order to address the issue of one team dominating in the training set, we use the all-but-K-majority where we consider the K most frequent teams for a payload under consideration. After trying out different values of K we select K = 3, which gives the best performance. For higher values of K, the pruning behaves like the baseline approach and for lower values it behaves like All-but-majority. On average each team gains about $40K$ samples in the training set as compared to all-but-majority pruning. Table~\ref{teams} shows the classifier performance. In this case also pruning performs better than baseline in 11 out of 20 teams, but as compared to all-but-majority the performance for most teams is better.\smallskip \\ 
\noindent\textbf{All-but-earliest (P-3):}
For this pruning we only retain the duplicates of the team that initiated the attack using a particular payload. This pruning technique retains all the non-deceptive duplicates while getting rid of the deceptive ones. Table~\ref{teams} shows the classifier performance. This pruning technique performs better than the baseline approach for 8 out of 20 teams. Comparing this result to all-but-majority (including all-but-K-majority) pruning indicates that deceptive duplicates are informative in attributing an attack to a team and should not be ignored completely.\smallskip \\
\noindent\textbf{All-but-most-recent (P-4):}
In this pruning we repeat a similar procedure like All-but-earliest but instead of retaining the duplicates of the team that initiated an attack, we retain the duplicates of the team that used the payload last in the training set. Since the data is sorted according to time, the last attacker becomes the most recent attacker for the test set. Table~\ref{teams} shows the classifier performance.

\begin{table}[h!]
	\caption{\textmd{Pruning technique performance comparison.}}
	\label{teams}
	\centering
	\tiny
	\renewcommand{\arraystretch}{1.5}
	
	\begin{tabular}{|p{0.7cm}|p{0.7cm}|p{0.7cm}|p{0.7cm}|p{0.7cm}|p{0.7cm}|} 
		\hline
		{\bf Teams} &  {\bf RF} &  {\bf P-1(RF)} &  {\bf P-2(RF)} &  {\bf P-3(RF)} &  {\bf P-4(RF)} \\ \hline 
	
		\textsf{T-1} & 0.45 & 0.16 & \textbf{0.46} & 0.15 & 0.15\\ 
		\textsf{T-2} & 0.22 & 0.28 & \textbf{0.30} & 0.15 & 0.14\\ 
		\textsf{T-3} & 0.30 & 0.53 & 0.29 & \textbf{0.57} & \textbf{0.57}\\ 
		\textsf{T-4} & 0.26 & \textbf{0.33} & 0.27 & 0.31 & 0.32\\ 
		\textsf{T-5} & 0.26 & 0.38 & \textbf{0.45} & 0.40 & 0.42\\ 
		\textsf{T-6} & \textbf{0.50}  & 0.27 & 0.24 & 0.31 & 0.26\\ 
		\textsf{T-7} & 0.45 & \textbf{0.59} & 0.58 & 0.19 & 0.49\\ 
		\textsf{T-8} & 0.42 & 0.52 & 0.52 & 0.51 & \textbf{0.55}\\ 
		\textsf{T-9} & 0.41 & 0.65 & \textbf{0.68} & 0.52 & 0.53\\ 
		\textsf{T-10} & 0.30 & 0.54 & 0.34 & 0.55 & \textbf{0.57}\\ 
		\textsf{T-11} & \textbf{0.37} & 0.27 & 0.35 & 0.27 & 0.29\\ 
		\textsf{T-12} & 0.24 & \textbf{0.37 }& \textbf{0.37} & 0.25 & 0.22\\ 
		\textsf{T-13} & 0.35 & 0.27 & \textbf{0.37} & 0.29 & 0.27\\ 
		\textsf{T-14} & \textbf{0.42} & 0.27 & 0.40 & 0.30 & 0.30\\ 
		\textsf{T-15} & \textbf{0.30} & 0.20 & 0.27 & 0.21 & 0.20\\ 
		\textsf{T-16} & \textbf{0.42} & 0.28 & 0.22 & 0.32 & 0.31\\ 
		\textsf{T-17} & 0.43 & \textbf{0.45} & 0.35 & 0.43 & 0.40\\ 
		\textsf{T-18} & \textbf{0.48} & 0.39 & 0.43 & 0.41 & 0.40\\ 
		\textsf{T-19} & 0.41 & \textbf{0.65} & 0.58 & 0.54 & 0.60\\ 
		\textsf{T-20} & \textbf{0.48} & 0.16 & 0.16 & 0.16 & 0.17\\ 
		\hline
	\end{tabular}
	\vspace{-1em}
	
\end{table}

Table~\ref{summary} gives the summary of the prediction results for all the pruning techniques in comparison with the random forest baseline approach. In the pruning techniques All-but-K-majority works best with an average accuracy of 0.42.

\begin{table}[h!]
	\caption{\textmd{Summary of Prediction results averaged across all Teams}}
	\label{summary}
	\centering
	\tiny
	\renewcommand{\arraystretch}{1.5}
	
	\begin{tabular}{|p{4cm}|p{2cm}|} 
		\hline
		{\bf Method} &  {\bf Average Performance} \\ \hline \hline
		
		\textsf{Baseline Approach (RF)} & 0.37\\ \hline
		\textsf{All-but-majority Pruning (RF)} & 0.40\\ \hline
		\textsf{All-but-K-majority Pruning (RF)} & \textbf{0.42}\\ \hline
		\textsf{All-but-earliest Pruning (RF)} & 0.34\\ \hline
		\textsf{All-but-most-recent Pruning (RF)} & 0.36\\ \hline
		
	\end{tabular}
	\vspace{-1em}
	
\end{table}

\section{Conclusion}
\label{con}
In this paper, we study cyber-attribution by examining DEFCON CTF data - which provides us with ground-truth on the culprit responsible for each attack. We frame cyber-attribution as a classification problem and examine it using several machine learning approaches. We find that deceptive incidents account for the vast majority of misclassified samples and introduce heuristic pruning techniques that alleviate this problem somewhat. Moving forward, we look to employ a
more principled approach to counter deception based on our previously established theoretical framework for reasoning about cyber-attribution~\cite{shakarian,shakarian14}. In particular we wish to employ temporal reasoning to tackle the problem of deceptive attacks. This opens up interesting research questions in particular identifying hacking group from a series of attacks over a period of time, differentiating between deceptive hacking groups in time series data. This is a knowledge engineering challenge which calls for development of efficient and scalable algorithms. 

\section{ACKNOWLEDGMENT}
Some of this work was supported by the U.S. Office of Naval Research and ASU Global Security Initiative (GSI).

\bibliographystyle{abbrv}

\bibliography{ref}

\end{document}